\documentclass{article}
\usepackage{spconf,amsmath,graphicx}
\usepackage{threeparttable}
\usepackage{float}
\usepackage{subfigure}
\usepackage{graphicx}
\usepackage{multirow}
\usepackage{booktabs}
\usepackage{tabularx}
\usepackage[hidelinks]{hyperref}
\usepackage{amssymb}
\usepackage{makecell}

\title{TEA-PSE 3.0: Tencent-Ethereal-Audio-Lab Personalized Speech Enhancement System for ICASSP 2023 DNS-Challenge}
\name{
\begin{tabular}{@{}c@{}}
Yukai Ju$^{1,2}$, Jun Chen$^{1}$, Shimin Zhang$^{2}$, Shulin He$^{1}$, Wei Rao$^{1}$, Weixin Zhu$^{1}$ \\ \textit{Yannan Wang$^{1}$, Tao Yu$^{1}$, Shidong Shang$^{1}$}
\end{tabular}
}
\address{$^1$ Tencent Ethereal Audio Lab, Tencent Corporation, Shenzhen, China \\
$^2$ Audio, Speech and Language Processing Group (ASLP@NPU), \\ Northwestern Polytechnical University, Xi'an, China}
\begin{document}
\ninept
\maketitle

\begin{abstract}
    This paper introduces the Unbeatable Team's submission to the ICASSP 2023 Deep Noise Suppression (DNS) Challenge. We expand our previous work, TEA-PSE, to its upgraded version -- TEA-PSE 3.0. Specifically, TEA-PSE 3.0 incorporates a residual LSTM after squeezed temporal convolution network (S-TCN) to enhance sequence modeling capabilities. Additionally, the local-global representation (LGR) structure is introduced to boost speaker information extraction, and multi-STFT resolution loss is used to effectively capture the time-frequency characteristics of the speech signals. Moreover, retraining methods are employed based on the freeze training strategy to fine-tune the system. According to the official results, TEA-PSE 3.0 ranks 1st in both ICASSP 2023 DNS-Challenge track 1 and track 2.

\end{abstract}
\vspace{-3pt}


\vspace{-15pt}
\section{Introduction}
\vspace{-7pt}
    
    Our previous work, the Tencent-Ethereal-Audio-Lab personalized speech enhancement (TEA-PSE)~\cite{ju2022tea}, ranked 1st in the ICASSP 2022 Deep Noise Suppression (DNS) Challenge. Building upon this success, we advance the previous model and propose our upgraded system, TEA-PSE 3.0, for this year's DNS Challenge. First, inspired by the derivative operator module in TaylorEnhancer~\cite{li2022general}, we introduce a residual LSTM after every squeezed temporal convolution network (S-TCN) layer to enhance the sequence modeling capability. Second, we utilize the local-global representation (LGR)~\cite{he2022local} structure to boost better speaker information extraction. Third, we adopt the multi-STFT resolution loss function~\cite{yamamoto2020parallel} to effectively capture the time-frequency characteristics of the speech signals. Finally, we leverage a more effective three-step training strategy. Specifically, we first train the stage-one model and freeze this model to train the stage-two model. Then we load this pre-trained two-stage model and fine-tune all trainable parameters with the second stage's loss function. Based on the final results, our model achieves 1st place in both headset and non-headset tracks~\cite{dubey2023icassp}.

\vspace{-7pt}
\section{Proposed method}

\vspace{-8pt}
\subsection{TEA-PSE 3.0 network}
\vspace{-3pt}

\begin{figure}[h]
        \centering
        \includegraphics[width=1\linewidth]{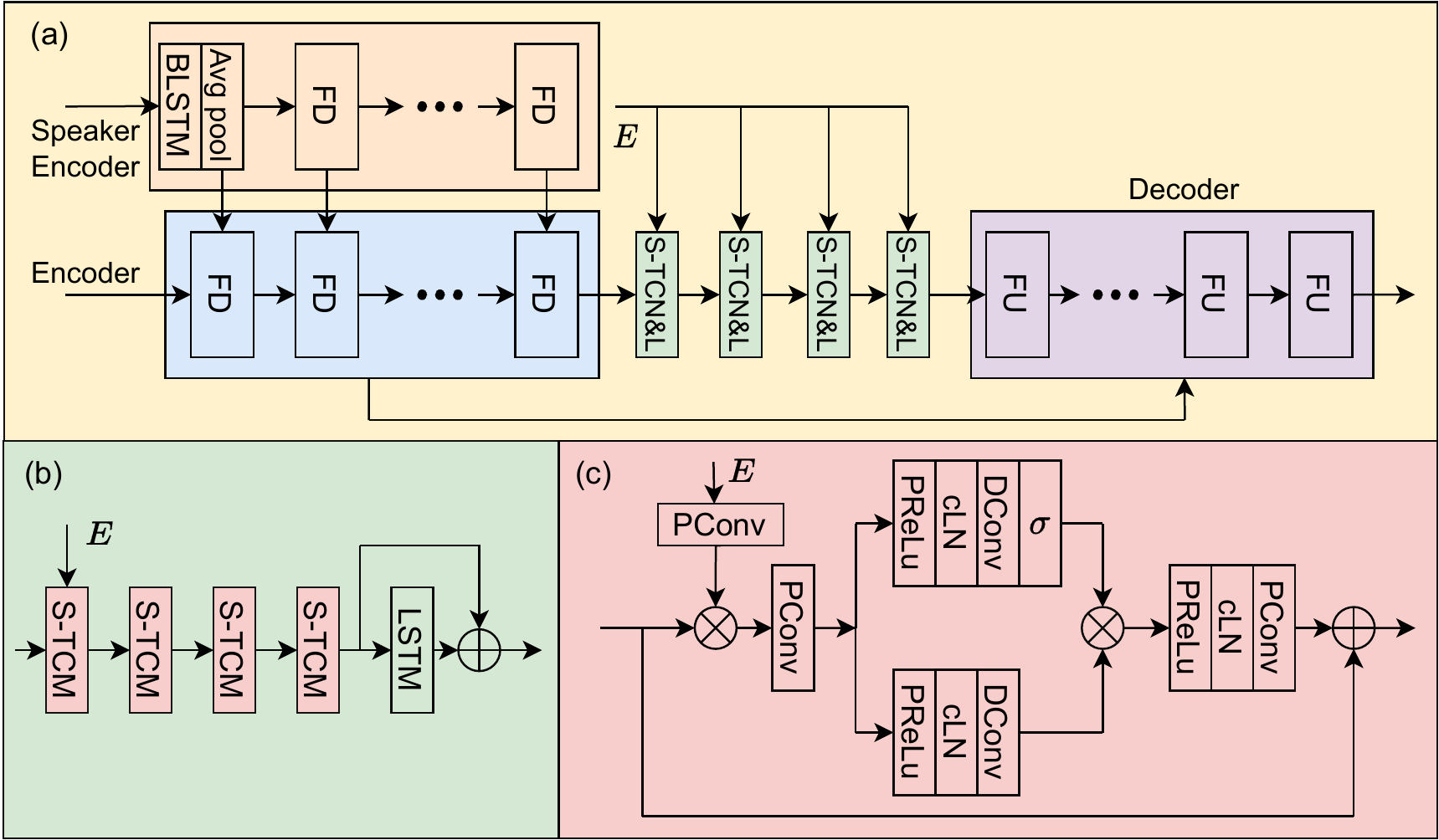}
        \vspace{-22pt}
        \caption{Details of (a) MAG-Net structure; (b) S-TCN\&L structure; (c) S-TCM structure.}
        \label{fig:overall}
        \vspace{-0.6cm}
    \end{figure}

    The proposed model maintains the two-stage framework of TEA-PSE~\cite{ju2022tea}, consisting of MAG-Net and COM-Net, to handle magnitude and complex-valued features, respectively. Fig.~\ref{fig:overall}(a) depicts the MAG-Net in detail, where $E$ represents the speaker embedding obtained from the pre-trained ECAPA-TDNN network.

    \textbf{Encoder and decoder.} The encoder is comprised of multiple frequency down-sampling (FD) layers, while the decoder is stacked by several frequency up-sampling (FU) layers. Each FD layer starts with a gated convolutional layer (GConv) to down-sample the input spectrum, followed by a cumulative layer norm (cLN) and PReLU. The FU layer is almost the same as the FD layer instead of replacing the GConv with transposed gated convolutional layer (TrGConv) to perform up-sampling.

    \textbf{Sequence modeling structure.} S-TCN consists of multiple squeezed temporal convolution modules (S-TCMs), as shown in Fig~\ref{fig:overall}(c). To further enhance the model's sequence modeling capabilities, we add a residual LSTM after every S-TCN module (termed as S-TCN$\&$L), as inspired by~\cite{li2022general}. Fig~\ref{fig:overall}(b) shows the modified S-TCN$\&$L structure. The speaker embedding is combined with the latent feature only in the first S-TCM layer of the S-TCN module using the multiply operation.

    \textbf{Local-global representation.} As local and global features of the speaker's enrollment speech (i.e. anchor) are both essential for target speaker extraction, we particularly incorporate the LGR structure~\cite{he2022local} into our model, as shown in Fig~\ref{fig:overall}(a). The speaker encoder consists of a bidirectional LSTM (BLSTM) and several FD layers, with the enrollment speech's magnitude as input. Note that there is an additional dense layer after the BLSTM to keep its dimension consistent with the input, and an average pooling operation is applied along the time dimension. The output of the speaker encoder is concatenated with the output of the previous FD layers in the encoder, corresponding to the further fusion of speaker information.

\vspace{-13pt}
\subsection{Loss function}
\vspace{-3pt}

    We employ several loss functions to train our model. Specifically, the scale-invariant signal-to-noise ratio (SI-SNR) loss $\mathcal{L}_{\text{si-snr}}$ and the power-law compressed phase-aware loss (magnitude loss $\mathcal{L}_\text{mag}$ and phase loss $\mathcal{L}_\text{pha}$) are used. Additionally, we use asymmetric loss $\mathcal{L}_{\text{asym}}$ to constrain the estimated spectrum to avoid over-suppression. These loss functions are defined in the same way as our previous work~\cite{ju2023tea}. First, we only train MAG-Net with $\mathcal{L}_{1}$. Following that, the pre-trained parameters of MAG-Net are frozen and only the COM-Net is optimized by $\mathcal{L}_{2}$. 
    \vspace{-8pt}
    \begin{equation}
    \begin{aligned}
    \mathcal{L}_{1}&=\mathcal{L}_\text{si-snr} + \frac{1}{M}\sum_{m=1}^{M}({\mathcal{L}_\text{mag} + \mathcal{L}_\text{asym}}),
    \\
    \mathcal{L}_{2}&=\mathcal{L}_{\text{si-snr}} + \frac{1}{M}\sum_{m=1}^{M}({\mathcal{L}_{\text{mag}} + \mathcal{L}_{\text{pha}} + \mathcal{L}_{\text{asym}}}).
    \end{aligned}
    \end{equation}
    \vspace{-2pt}
    Furthermore, for all frequency domain loss functions, we explore multi-STFT resolution~\cite{yamamoto2020parallel}, where $m$ indicates the scale corresponding to different STFT configurations. We train MAG-Net and COM-Net sequentially as described above and then load these pre-trained models to retrain the entire system using $\mathcal{L}_{2}$.

\vspace{-12pt}
\section{Experiments} 

\vspace{-9pt}
\subsection{Dataset}
\vspace{-3pt}

    We use the ICASSP 2022 DNS-challenge full-band dataset~\cite{dubey2022icassp} for experiments. The noise data originates from DEMAND, Freesound, and AudioSet. We generate 100,000 room impulse responses (RIRs) based on the image method~\cite{allen1979image} with $\text{RT60}\in [0.1, 1.0]$s.

\vspace{-9pt}
\subsection{Training setup}
\vspace{-3pt}

    The window length and frameshift are 20ms and 10ms, respectively. For multi-STFT resolution loss, we use 3 different groups with FFT length $\in\{512,1024,2048\}$, window length $\in\{480,960,1920\}$, and frameshift $\in\{240,480,960\}$. We use FFT length $1024$, window length $960$, and frameshift $480$ for single-STFT resolution loss. The Adam optimizer is used to optimize our models, and the initial learning rate is $1e^{-3}$. The learning rate will be halved if the validation loss has no decrease for 2 epochs.  We use on-the-fly data generation to increase the diversity of generated data and save storage space, which keeps the same settings as TEA-PSE.

    The encoder and decoder consist of 6 FD layers and 6 FU layers, respectively. The GConv and TrGConv in both the encoder and decoder have a kernel size and stride of (2, 3) and (1, 2) in the time and frequency axis, respectively. All GConv and TrGConv layers' channels are set to 64. The S-TCN$\&$L module has 4 S-TCM layers with a kernel size of 5 for dilated Conv (DConv) and a dilation rate of $\{1,2,5,9\}$, respectively, and a hidden size of 512 for the LSTM. All convolution channels in S-TCN$\&$L are set to 64 except for the last pointwise Conv (PConv) layer. We stack 4 S-TCN$\&$L groups to establish long-term relationships between consecutive frames and combine speaker embeddings. For the speaker encoder, we use a BLSTM with a hidden size of 512 and 5 FD layers, and all GConv layers' channels in the speaker encoder are set to 1.

\vspace{-10pt}
\subsection{Results and analysis} 
\vspace{-5pt}

    According to the blind test set results in Table~\ref{tab:blind}, several observations can be made. First, adding a residual LSTM after every S-TCN module improves performance. Second, the LGR structure has proven to be effective in boosting speaker information extraction. Third, by using the multi-STFT resolution loss function, the proposed method achieves a significant improvement of 0.015 and 0.042 in OVRL for track 1 and track 2, respectively. Finally, retraining the dual-stage network with the pre-trained model provides an additional gain in performance. 
    
    Table~\ref{tab:blind_mos} shows the mean opinion score (MOS) and word accuracy (WAcc) results on the DNS 2023 blind test set. TEA-PSE 3.0 has the highest BAK and OVRL. Besides, compared with unprocessed speech, the SIG and WAcc of the submission model are decreased, which is reasonable since the model introduces slight distortion to the extracted speech.
    
    TEA-PSE 3.0 has a total of 22.24 million trainable parameters. The number of multiply-accumulate operations (MAC) of TEA-PSE 3.0 is 19.66G per second. The average real-time factor (RTF) per frame for the submission method exported by ONNX is 0.46 on an Intel(R) Xeon(R) CPU E5-2678 v3 clocked at 2.4 GHz. 

    \begin{table}[h]
    \centering
    \setlength{\tabcolsep}{4.5pt}
    \vspace{-16pt}
    \caption{PDNSMOS P.835 results on the DNS 2023 blind test set.}
    \footnotesize
    \vspace{2pt}
    \begin{tabular}{@{}llcccccc@{}}
    \toprule
    ID & Method & \multicolumn{3}{c}{Track 1} & \multicolumn{3}{c}{Track 2} \\
    \midrule
    & & SIG & BAK & OVRL & SIG & BAK & OVRL \\
    \midrule
    1 & Noisy & \textbf{4.152} & 2.369 & 2.709 & \textbf{4.046} & 2.159 & 2.497 \\
    2 & TEA-PSE & 3.996 & 4.010 & 3.528 & 3.850 & 3.884 & 3.342 \\
    3 & ~ + LSTM & 4.044 & 4.032 & 3.562 & 3.910 & 3.915 & 3.384 \\
    4 & ~ ~ + LGR & 4.083 & 4.050 & 3.603 & 3.949 & 3.932 & 3.429 \\
    5 & ~ ~ ~ + multi-STFT & 4.086 & \textbf{4.058} & 3.618 & 3.986 & 3.949 & 3.471 \\
    6 & ~ ~ ~ ~ + retrain & 4.108 & 4.053 & \textbf{3.645} & 3.993 & \textbf{3.951} & \textbf{3.493} \\
    \toprule
    \end{tabular}
    \label{tab:blind}
    \end{table}

    \begin{table}[h]
    \centering
    \setlength{\tabcolsep}{3.8pt}
    \vspace{-31pt}
    \caption{MOS and WAcc results on the DNS 2023 blind test set.}
    \footnotesize
    \vspace{2pt}
    \begin{tabular}{@{}lcccccccc@{}}
    \toprule
    System & \multicolumn{4}{c}{Track 1} & \multicolumn{4}{c}{Track 2} \\
    \midrule
    & SIG & BAK & OVRL & WAcc & SIG & BAK & OVRL & WAcc \\
    \midrule
    Noisy & \textbf{3.76} & 1.22 & 1.22 & \textbf{0.843} & \textbf{3.83} & 1.22 & 1.24 & \textbf{0.857} \\
    DNS Baseline & 3.20 & 2.67 & 2.34 & 0.687 & 3.22 & 2.68 & 2.38 & 0.727 \\
    TEA-PSE 3.0 & 3.52 & \textbf{2.88} & \textbf{2.71} & 0.761 & 3.64 & \textbf{2.92} & \textbf{2.72} & 0.768 \\
    \toprule
    \end{tabular}
    \label{tab:blind_mos}
    \end{table}

\vspace{-25pt}
\section{Conclusions}
\vspace{-7pt}

    The proposed TEA-PSE 3.0 utilizes the S-TCN$\&$L module, which provides enhanced sequence modeling capabilities. With the LGR structure, our method can make better use of speaker information. Additionally, we investigate the effectiveness of the multi-STFT resolution loss function, comparing it with the single-STFT resolution. Based on the freeze training strategy, we explore the effect of model retraining. According to the official challenge results, TEA-PSE 3.0 ranks 1st in both tracks.



\footnotesize
\bibliographystyle{IEEE}

\bibliography{strings}
\end{document}